\pdfoutput=1
\documentclass[11pt]{article}

\usepackage[utf8]{inputenc}
\usepackage[T1]{fontenc}
\usepackage{lmodern}
\usepackage{microtype}
\usepackage[margin=1in]{geometry}
\usepackage{amsmath}
\usepackage{booktabs}
\usepackage{tabularx}
\usepackage{xurl}
\usepackage[hidelinks]{hyperref}

\hypersetup{
  pdftitle={Data Centers: The Error of Generalizing Environmental and Socioeconomic Impacts},
  pdfauthor={Elzo Brito dos Santos Filho}
}

\newcommand{\LAeq}{\ensuremath{L_{\mathrm{Aeq}}}}
\newcommand{\cubicm}{m\textsuperscript{3}}

\title{Data Centers: The Error of Generalizing Environmental and Socioeconomic Impacts\\[0.6em]
  \large The Need for a Contextualized Assessment of Computing Infrastructure in Brazil and Worldwide}
\author{Elzo Brito dos Santos Filho}
\date{November 2026}

\begin{document}

\maketitle

\begin{abstract}
The rapid expansion of artificial intelligence and cloud computing has intensified debates about the environmental and socioeconomic impacts of data centers, including electricity and water consumption, noise, land use, and employment. A recurring methodological problem in these debates is the use of global data, or of evidence produced in a particular country, to draw conclusions about facilities located in very different settings. This paper argues that the impacts of a data center depend on the geographic, climatic, energy, technological, economic, and regulatory conditions of the territory where it operates, and must therefore be assessed at the appropriate territorial and technological scale. Using Brazil as a case, we discuss the country's predominantly renewable electricity mix and the curtailment of renewable generation, the reported prevalence of closed-loop cooling, national and municipal noise regulations, the water-efficiency requirement of the new Redata tax regime, and the ongoing litigation over the Pecém data center in the state of Ceará. We also examine a recent study of data centers in Northern Virginia, showing the limits of its external validity and of its noise measurements. We propose a multiscale assessment framework and conclude that international evidence should inform which questions are asked, not predetermine the answers---a principle that applies equally to claims of harm and to claims of safety.
\end{abstract}


\section{Introduction}

The expansion of artificial intelligence and cloud computing has driven a significant increase in demand for computing infrastructure. In parallel, discussions about the environmental and socioeconomic impacts of data centers have intensified, particularly regarding water consumption, energy demand, noise pollution, land occupation, and job creation. However, there is a recurring problem in how these arguments are constructed: global data, or studies conducted in particular countries, are used to support conclusions about facilities located in entirely different settings.

A data center is not an infrastructure independent of the environment in which it is embedded. Its operation depends on the geographic, climatic, energy, technological, economic, and regulatory characteristics of the territory where it is built. Therefore, we cannot use the environmental impacts identified in U.S. facilities as sufficient evidence of the impacts a Brazilian facility will produce. Likewise, we cannot indiscriminately use data from São Paulo to assess a project located in Ceará, since, even within a single country, there are significant differences among regions, states, and municipalities.

This does not mean disregarding international research, but rather understanding the limits of its external validity. Global data are important for characterizing trends, while regional studies make it possible to identify specific phenomena. Neither, however, replaces an assessment of the concrete conditions under which a given project will be installed.

\section{Electricity Consumption and the Characteristics of the Power Mix}

One of the main arguments in discussions about the environmental impacts of data centers concerns their high electricity consumption. This consumption is often directly associated with increased greenhouse gas emissions, disregarding the fact that the carbon intensity of electricity generation varies substantially across countries.

According to the 2026 Brazilian Energy Balance (\emph{Balanço Energético Nacional}, BEN), published by the Energy Research Office (\emph{Empresa de Pesquisa Energética}, EPE), renewable sources accounted for 86.8\% of Brazil's electricity mix in 2025~\cite{epe2026}. This characteristic distinguishes Brazil from countries whose electricity generation depends predominantly on fossil fuels. Two data centers with identical equipment, the same computing capacity, and exactly the same electricity consumption may have significantly different carbon footprints depending on the energy sources they use.

Moreover, Brazil has a particularity that must be incorporated into the discussion: the existence of surplus renewable generation in certain periods and regions. The growth of wind and solar generation has led to situations in which part of the potentially available energy must be restricted, a phenomenon known as curtailment. The National Electric System Operator (\emph{Operador Nacional do Sistema Elétrico}, ONS) itself, in its PAR/PEL 2025 plan, identifies data centers as a promising alternative for absorbing part of these surpluses and potentially reducing generation cuts~\cite{ons2025}.

In this context, while some countries need to expand their electricity generation to meet growing computing demand, Brazil has opportunities to turn energy surpluses into processing capacity, digital services, and economic development. Naturally, the existence of these surpluses does not eliminate the need for investment in transmission or for technical grid-connection studies. It does show, however, that the relationship between computing expansion and energy availability cannot be treated as a universal problem, independent of the characteristics of the local power system.

\section{Water Consumption and Cooling Technologies}
\label{sec:water}

Water consumption is another example of how generalizing data can lead to mistaken interpretations. When it is claimed that data centers consume large amounts of water, sufficient information is rarely provided about the cooling technologies used by the facilities from which the estimates were derived.

There are different thermal management architectures. Some use evaporative systems, which consume water continuously to reject heat. Others use closed loops combined with dry heat rejection, which allow the coolant to be recirculated and significantly reduce the need for make-up water. It is therefore not technically appropriate to treat these architectures as equivalent.

In 2025, a survey conducted by Brasscom, the Brazilian association of information and communication technology companies, with support from the Brazilian Data Center Association (\emph{Associação Brasileira de Data Centers}, ABDC) and technical review by FADURPE, estimated that approximately 80\% of Brazil's data center stock used closed-loop cooling systems, a share projected to reach 90\% by 2030~\cite{brasscom2025}. Although these figures come from an industry study and need to be complemented by independent measurements, they show the importance of understanding the technological characteristics of Brazilian infrastructure before applying international estimates of water consumption.

Brazil's geographic conditions themselves influence engineering decisions. States such as Ceará and Rio de Janeiro have strategic advantages related to their proximity to submarine cable landing points, which are essential infrastructure for international communication. However, installing equipment in coastal regions requires additional care regarding salinity, humidity, and corrosion caused by salt-laden sea air.

In September 2025, a report by the newspaper \emph{Folha de S.Paulo} found that none of the data center operators consulted in Ceará reported using evaporative cooling~\cite{folha2025}. The report also highlighted the need for special treatments to protect outdoor equipment against salt-laden air. These are examples of environmental constraints that shape technological choices and show why Brazilian facilities should not be presumed equivalent to those in other countries. This information, however, was self-reported by the companies, and the largest data center under construction in the state shows why such statements are no substitute for independent verification.

The Pecém Data Center, in Caucaia, has a planned capacity of 300~MW, an area of about 70 hectares, 120 diesel generators, and groundwater abstraction for cooling~\cite{mpf2026}. The company responsible for the project states that it uses closed-loop cooling with minimal make-up water and consumption below 30~\cubicm{} per day, and maintains, based on what it describes as independent hydrogeological studies, that the abstraction does not compromise neighboring communities~\cite{dn2026}. The Federal Prosecution Service (\emph{Ministério Público Federal}, MPF) and the Federal Public Defender's Office (\emph{Defensoria Pública da União}, DPU), in turn, have pointed to a discrepancy between the water volumes reported in the Simplified Environmental Report (\emph{Relatório Ambiental Simplificado}, RAS) and in the installation license~\cite{dn2026}, and an MPF technical report, reported in the press in December 2025, mentions a consumption of 88~\cubicm{} per day~\cite{istoe2025}. In September 2026, the two institutions filed a public civil action to prevent the facility from starting operations before an environmental impact study and report (EIA/RIMA) is prepared, pumping tests are carried out with validation of the hydrogeological model, and the Anacé people are given free, prior, and informed consultation, arguing that the surrounding communities depend mostly on wells in a region with a history of water scarcity~\cite{mpf2026}. Ceará's state environmental agency (\emph{Superintendência Estadual do Meio Ambiente}, Semace) maintains that the licensing was sufficient, arguing that the area had already been covered by an EIA/RIMA prepared for the Hydrogen Hub~\cite{dn2026}.

The case also shows both the reach and the limits of aggregate indicators. Law No.~15,504/2026, which established the Special Tax Regime for Data Center Services (Redata), requires participating companies to achieve a water usage effectiveness (WUE) of 0.05~L/kWh or less, measured annually~\cite{brasil2026}. According to DatacenterDynamics, this threshold is stricter than the global values reported by major cloud providers for 2025: 0.12~L/kWh for AWS and 0.27~L/kWh for Microsoft~\cite{dcd2026}. As an order of magnitude, assuming 300~MW of IT load in continuous operation, even the figure of 88~\cubicm{} per day would correspond to about 0.012~L/kWh, well below the limit;\footnote{$88{,}000$~L/day $\div$ ($300{,}000$~kW $\times$ 24~h/day) $\approx 0.012$~L/kWh. Because WUE is computed on the IT energy actually consumed, a lower average load yields a proportionally higher value.} with a lower load, the index increases proportionally. Even so, the controversy persists, because the relevant question is not only whether the facility uses little water relative to the energy it consumes, but whether the aquifer can sustain the abstraction without compromising priority uses, such as human consumption. Furthermore, the indicator measures the volume used, not its source, and participation in the regime is optional~\cite{dcd2026}.

It is also necessary to distinguish an internal closed loop from a system whose final heat rejection takes place without evaporation. A closed loop may be coupled to an external evaporative cooling tower, and even non-evaporative systems require water for initial filling, loss make-up, and auxiliary uses. For this reason, the assessment of water consumption must consider the facility's entire thermal architecture, not merely the label attached to its cooling system.

Therefore, just as foreign research does not prove that a Brazilian data center will cause harm, national averages, efficiency indicators, and corporate statements do not prove that it will not. Before claiming that a given facility does, or does not, threaten water supply, we must know its cooling architecture, its actually measured consumption, the source of its water (surface water, groundwater, or reclaimed water), the water availability and competing uses in the basin or aquifer, and its cumulative impacts together with other projects in the region.

\section{Noise Pollution and the Particularities of Legislation}

Noise from data center equipment is a physical consequence of its operation. Ventilation systems, compressors, transformers, chillers, and generators produce different sound pressure levels. However, the existence of a sound source is not sufficient to automatically characterize a situation of harmful environmental pollution.

In Brazil, there are regulatory instruments governing noise emissions from industrial and commercial activities. Resolution No.~1/1990 of the National Environment Council (\emph{Conselho Nacional do Meio Ambiente}, CONAMA) sets criteria for noise emissions~\cite{conama1990}, while the Brazilian standard ABNT NBR 10151:2019 establishes technical parameters for assessing sound pressure levels in inhabited areas~\cite{abnt10151}.

For strictly residential areas, hospitals, and schools, the standard sets reference levels of 50~dB(A) during the daytime and 45~dB(A) at night. In addition, municipalities may establish additional limits according to their urban characteristics. The city of São Paulo, for example, has its own legislation setting nuisance parameters associated with urban zoning and the operating hours of economic activities~\cite{saopaulo2016}.

Therefore, assessing the acoustic impacts of a data center must take into account national legislation, applicable state regulations, municipal limits, the zoning classification of the site, and the location of nearby communities. It must also consider the engineering solutions used to reduce sound propagation, such as equipment enclosures, acoustic barriers, and the appropriate placement of cooling systems.

The existence of regulations does not automatically guarantee compliance, but it provides objective criteria for investigating possible irregularities. We cannot conclude that a given Brazilian facility will produce noise pollution simply because researchers have identified acoustic problems at foreign facilities.

\section{Electrical Infrastructure, Tariffs, and Air Pollution}

The possibility of overloading the electrical infrastructure, supply interruptions, harmonic distortion, and increases in residential tariffs must also be analyzed in light of the particularities of each country's energy system.

Brazil has its own structure for the generation, transmission, distribution, and trading of electricity. Large consumers may operate under different contracting arrangements, use high-voltage connections, and require dedicated electrical infrastructure. There are also technical and regulatory requirements regarding power quality, established by the National Electric Energy Agency (\emph{Agência Nacional de Energia Elétrica}, ANEEL) and by the ONS~\cite{aneel_prodist}.

A projection of residential tariff increases made in the United States is conditioned by the utilities' remuneration model, the allocation of infrastructure expansion costs, and the regulation of that market. Transposing such a projection to Brazil without investigating the differences between the two systems is a methodologically inappropriate extrapolation.

The same distinction must be made regarding air pollution. The high share of renewable sources in Brazil's electricity mix reduces the average emission intensity associated with electricity consumption, while any local emissions from emergency generators must be assessed separately, considering their frequency of use, operating duration, and environmental control mechanisms.

It is important to note that emergency generators are also used in hospitals, industrial plants, shopping centers, and other critical infrastructure. Their presence is not an environmental feature exclusive to data centers. To demonstrate a particularly relevant impact, it would be necessary to assess actual emissions and establish appropriate comparative parameters.

\section{Land Occupation and Environmental Licensing}

Another frequently raised argument concerns the loss of green areas and the transformation of urban space resulting from data center construction. However, land occupation is a characteristic common to the installation of virtually any large-scale project, whether an industrial plant, a hospital, a logistics center, a university, or a computing facility.

The question should not simply be whether a data center occupies a physical area; rather, one should investigate the characteristics of the site, its previous use, the municipal zoning, the need for vegetation clearance, and the environmental impacts of construction.

In Brazil, there are legal instruments for urban planning, environmental protection, and the licensing of economic activities. A facility built in an industrial area previously designated for this type of activity presents substantially different conditions from one that requires clearing native vegetation or is located in an environmentally sensitive region.

To characterize data center land occupation as a specific problem, it would be necessary to show that its impacts are proportionally greater than those of other comparable economic activities. The mere existence of a building is not sufficient evidence to support that conclusion.

\section{Job Creation and the Economic Role of Computing Infrastructure}

The number of jobs generated by data centers must also be analyzed in light of the economic nature of these facilities and the different phases of their life cycle.

During design, construction, and deployment, there is significant demand for professionals in civil engineering, architecture, electrical installations, telecommunications, automation, refrigeration, security, and many other technical specialties. Once the facility enters operation, the workforce profile changes, with a greater share of professionals specialized in critical infrastructure, information security, networking, cloud computing, systems administration, and operational reliability.

A study released by FGV Projetos, the consulting arm of the Getulio Vargas Foundation (FGV), in July 2026 estimates that deploying a 100~MW data center dedicated to artificial intelligence may mobilize approximately R\$~25 billion in investment, generate about 12,560 direct and indirect jobs during construction and deployment, and sustain approximately 1,860 permanent positions~\cite{fgv2026}. The study, commissioned by Scala Data Centers and Norgás, relies on economic projections based on an input--output matrix, and its results should be interpreted in accordance with that methodology.

Moreover, operating modern computing services does not require all professionals to be physically present where the equipment is installed. Activities related to monitoring, systems administration, information security, and cloud service management can be performed remotely, distributing job opportunities across different cities, states, and even countries.

Assessing a data center's economic contribution solely by the number of workers present at its premises reflects a limited understanding of its purpose. These facilities are capital- and technology-intensive, and their main function is to provide computing capacity for other economic activities.

A data center should be evaluated not only by the jobs it creates directly, but also by the jobs its infrastructure makes possible. Software companies, financial institutions, universities, hospitals, educational platforms, research organizations, and countless other activities depend on the availability of computing services to operate, grow, and develop new products.

\section{Digital Sovereignty, the LGPD, and Distributed Economic Benefits}

The benefits produced by computing infrastructure are not necessarily concentrated in the communities near its facilities. A data center built in a given municipality may serve millions of users throughout the country, in addition to providing services to foreign organizations.

In Brazil's case, there is an additional strategic dimension: the availability of infrastructure for storing and processing data within the national territory. Expanding this capacity can reduce dependence on foreign computing services, facilitate data residency policies, and strengthen mechanisms of governance, information security, and technological autonomy.

Brazil's General Personal Data Protection Law (\emph{Lei Geral de Proteção de Dados Pessoais}, LGPD) sets requirements for the processing of personal data and governs the conditions for its international transfer~\cite{brasil2018,anpd2024}. Although the law does not, as a general rule, require all Brazilian data to be stored exclusively in the country, the availability of domestic infrastructure can facilitate compliance with specific data residency requirements and reduce the need for certain international transfers.

Installing a data center in Brazil does not automatically guarantee LGPD compliance, but it broadens the possibilities for Brazilian organizations to exercise greater control over their data storage, processing, and governance strategies.

Furthermore, the existence of computing capacity within the national territory has implications for the development of artificial intelligence, scientific research, education, public administration, and business competitiveness. These are benefits that can hardly be measured solely by employment or tax revenue indicators in the municipality where the infrastructure was built.

\section{A Counterpoint to International Evidence}
\label{sec:counterpoint}

As a counterpoint to this contextualized perspective, we can examine the paper \emph{The Cloud Next Door: Investigating the Environmental and Socioeconomic Strain of Datacenters on Local Communities}, published by Ngata et~al.\ at the ACM COMPASS 2025 conference~\cite{ngata2025}.

The study investigates the environmental and socioeconomic effects of data centers on communities in Northern Virginia, United States, addressing issues related to natural resource consumption, power quality, tariffs, noise and air pollution, and the distribution of economic benefits.

The research has merit in drawing attention to the local impacts of digital infrastructure, which are often overlooked in aggregate analyses. However, its preliminary results are tied to the specific characteristics of the region studied and do not constitute sufficient evidence that the same phenomena will occur in countries with different environmental, technological, and regulatory conditions.

Regarding noise pollution, for example, the researchers report 22.3~dB in a control neighborhood about 3~km from a data center and 28.0~dB in another neighborhood approximately 60~m from a facility, both in Ashburn. These figures, however, cannot be compared with the limits of ABNT NBR 10151, either to suggest compliance or to indicate a violation. They are time-weighted average (TWA) values obtained with the Sound Level Meter smartphone app of the National Institute for Occupational Safety and Health (NIOSH), and the study does not report the frequency weighting, duration, time of day, number of measurements, or procedure used. TWA is an occupational exposure metric referenced to an eight-hour shift; for shorter measurements, it is lower than the A-weighted equivalent continuous sound level (\LAeq) of the measured period~\cite{niosh_understand}. With the 3~dB exchange rate adopted by NIOSH, for example, a 15-minute reading would yield a TWA about 15~dB below the corresponding \LAeq.\footnote{Assuming no threshold level, a 3~dB exchange rate, and an 8-hour (480-minute) criterion duration, $\mathrm{TWA} = L_{\mathrm{Aeq},T} + 10\log_{10}(T/480)$, where $T$ is the measurement duration in minutes and $L_{\mathrm{Aeq},T}$ is the equivalent level over that period. For $T = 15$, the correction is $10\log_{10}(15/480) \approx -15$~dB.} For the same reason, even the 5.7~dB difference between the two sites depends on both readings having had the same duration, which the study does not report. The NBR 10151 limits, in turn, are expressed as A-weighted equivalent levels obtained through specific procedures, which include distinguishing total, specific, and residual sound and assessing tonal and impulsive components~\cite{abnt10151}.

Moreover, NIOSH itself states that the app is not intended for compliance assessments~\cite{niosh2017}, and the study describes the noise sometimes as a low-frequency hum and sometimes as high-frequency sound associated with power distribution infrastructure, without presenting a frequency analysis that would allow the sources to be distinguished. The limitation, therefore, lies not only in transposing the results to Brazil: two readings without a documented protocol are insufficient to characterize noise pollution even in the region studied. In Brazil, an adequate assessment requires measurements in accordance with NBR 10151 and the applicable municipal legislation, preferably before and after the facility enters operation, with attention to tonal and low-frequency components. Indeed, semiannual disclosure of noise levels is among the requests in the public civil action concerning the Pecém Data Center, discussed in Section~\ref{sec:water}~\cite{mpf2026}.

Likewise, projections of residential tariff increases must take into account the characteristics of the U.S. electricity market, while discussions of water consumption must be confronted with the cooling architectures actually used in Brazilian facilities.

Concerns about community health must also be distinguished from the scientific demonstration of harm. The study does not present its own epidemiological investigation capable of establishing a causal relationship between the operation of the facilities analyzed and specific diseases. Its results may support new research hypotheses, but they do not prove that similar facilities will necessarily produce the same effects on other populations.

The authors themselves acknowledge the need to extend the investigation to other locations and data center configurations. The criticism, therefore, is not directed at conducting research on the local impacts observed in Virginia, but at the possible use of its conclusions as universal arguments against the installation of computing infrastructure in other territories.

\section{A Model for Contextualized Assessment}

A technically sound assessment of the socio-environmental impacts of data centers should consider different territorial scales and technological dimensions, as summarized in Table~\ref{tab:framework}.

\begin{table}[htbp]
  \centering
  \caption{Dimensions of a contextualized assessment of data center impacts.}
  \label{tab:framework}
  \begin{tabularx}{\linewidth}{@{}lX@{}}
    \toprule
    \textbf{Dimension} & \textbf{Elements to be considered} \\
    \midrule
    Country      & Electricity mix, national legislation, economic model, data protection, and digital sovereignty. \\
    Region       & Climate, water availability, energy generation, and connectivity infrastructure. \\
    State        & Environmental licensing, territorial planning, taxation, and infrastructure. \\
    Municipality & Urban zoning, noise limits, availability of public services, and tax revenue. \\
    Site         & Proximity to communities, site characteristics, and environmental conditions. \\
    Technology   & Cooling architecture, energy efficiency, computing capacity, and mitigation systems. \\
    Life cycle   & Construction, deployment, operation, modernization, and decommissioning. \\
    Economy      & Direct and indirect jobs, investment, productivity, and the economic activities enabled. \\
    Governance   & Oversight, transparency, auditing, and compliance with legal obligations. \\
    \bottomrule
  \end{tabularx}
\end{table}

It would also be necessary to develop comparative indicators that allow data centers to be assessed relative to other economic activities. Absolute figures for water consumption, electricity use, or land occupation have limited analytical value when presented without considering the facility's productive capacity, the economic value it generates, and the social benefits it provides.

International references remain important, but they should serve to identify risks, formulate hypotheses, develop methodologies, and compare different operating conditions. Demonstrating impacts requires evidence representative of the territory and of the technological architecture of the facility under investigation.

\section{Concluding Remarks}

The debate on the environmental and socioeconomic impacts of data centers needs to abandon the premise that these facilities produce uniform consequences regardless of country, region, or technology. The characteristics of the electricity mix, water availability, climatic conditions, environmental legislation, urban planning, tariff models, and the very architecture of the facilities directly influence the observed outcomes.

Brazil has particularities that must be part of any assessment of the expansion of its computing infrastructure. These include the high share of renewable sources in the electricity mix, the energy surpluses that exist in certain periods, the estimated predominance of closed-loop cooling systems, environmental and urban regulations, the opportunities associated with international connectivity, and the need to strengthen technological autonomy and the governance of national data.

These characteristics do not mean that Brazilian data centers are immune to environmental impacts, but they show that such impacts must be investigated on the basis of Brazilian evidence. Likewise, aggregate national data are not sufficient to determine the consequences of a specific facility without considering the particularities of the region, the state, the municipality, and the facility itself.

We cannot use global data or evidence produced in other countries as standalone proof of the problems a Brazilian community will face. We can use them to identify what needs to be investigated, which risks need to be mitigated, and which indicators should be monitored, but not to establish the conclusions of that investigation in advance.

A scientifically responsible analysis must also consider the economic and strategic benefits of computing infrastructure, recognizing that its effects extend beyond municipal boundaries and potentially contribute to the development of different sectors of society.

The central question, therefore, is not whether data centers are, by definition, harmful or beneficial, but under which technological, environmental, economic, and regulatory conditions these facilities can produce social benefits that exceed their costs.

The principles of engineering are universal, but the environmental and socioeconomic impacts of their application are contextual. Research conducted in another country can teach us which questions to ask in Brazil, but it cannot determine in advance which answers we will find.


\begin{thebibliography}{99}

\bibitem{aneel_prodist}
Agência Nacional de Energia Elétrica (ANEEL).
\emph{Procedimentos de Distribuição de Energia Elétrica no Sistema Elétrico Nacional -- PRODIST} [Electricity distribution procedures in the national electric system].
Available at: \url{https://www.gov.br/aneel/pt-br/centrais-de-conteudos/procedimentos-regulatorios/prodist}.

\bibitem{abnt10151}
Associação Brasileira de Normas Técnicas (ABNT).
\emph{ABNT NBR 10151:2019 -- Acústica: medição e avaliação de níveis de pressão sonora em áreas habitadas -- aplicação de uso geral} [Acoustics: measurement and assessment of sound pressure levels in inhabited areas -- general application].
Corrected version, 2020.

\bibitem{anpd2024}
Autoridade Nacional de Proteção de Dados (ANPD).
Resolução CD/ANPD nº~19, de 23 de agosto de 2024: Regulamento de Transferência Internacional de Dados [Resolution No.~19 of August 23, 2024: Regulation on international data transfer].
2024.

\bibitem{brasil2018}
Brasil.
Lei nº~13.709, de 14 de agosto de 2018: Lei Geral de Proteção de Dados Pessoais (LGPD) [Law No.~13,709 of August 14, 2018: General Personal Data Protection Law].
Available at: \url{https://www.planalto.gov.br/ccivil_03/_ato2015-2018/2018/lei/l13709compilado.htm}.

\bibitem{brasil2026}
Brasil.
Lei nº~15.504, de 15 de setembro de 2026: institui o Regime Especial de Tributação para Serviços de Datacenter (Redata) [Law No.~15,504 of September 15, 2026: establishes the Special Tax Regime for Data Center Services].
Available at: \url{https://www.planalto.gov.br/ccivil_03/_Ato2023-2026/2026/Lei/l15504.htm}.
Accessed: Nov.~25, 2026.

\bibitem{brasscom2025}
Brasscom.
\emph{Consumo de Energia e Água em Data Centers no Brasil} [Energy and water consumption in data centers in Brazil].
2025.
Available at: \url{https://brasscom.org.br/pdfs/consumo-de-energia-e-agua-em-data-centers/}.

\bibitem{conama1990}
Conselho Nacional do Meio Ambiente (CONAMA).
Resolução nº~1, de 8 de março de 1990: critérios e padrões de emissão de ruídos [Resolution No.~1 of March 8, 1990: noise emission criteria and standards].
1990.

\bibitem{dcd2026}
DatacenterDynamics.
Boom dos data centers coloca água de reúso no centro da infraestrutura para inteligência artificial [Data center boom puts reclaimed water at the center of AI infrastructure].
2026.
Available at: \url{https://www.datacenterdynamics.com/br/not%C3%ADcias/boom-dos-data-centers-coloca-%C3%A1gua-de-re%C3%BAso-no-centro-da-infraestrutura-para-intelig%C3%AAncia-artificial/}.
Accessed: Nov.~25, 2026.

\bibitem{dn2026}
Diário do Nordeste.
MPF e DPU recomendam adequações ambientais para início de atividades no data center do TikTok [MPF and DPU recommend environmental adjustments before the TikTok data center begins operations].
May~20, 2026.
Available at: \url{https://diariodonordeste.verdesmares.com.br/negocios/mpf-e-dpu-recomendam-adequacoes-ambientais-para-inicio-de-atividades-no-data-center-do-tiktok-1.3765405}.
Accessed: Nov.~25, 2026.

\bibitem{epe2026}
Empresa de Pesquisa Energética (EPE).
\emph{Balanço Energético Nacional 2026: ano-base 2025} [Brazilian Energy Balance 2026: base year 2025].
2026.
Available at: \url{https://www.epe.gov.br/}.

\bibitem{folha2025}
Folha de S.Paulo.
Cabos submarinos e energia limpa transformam Ceará em polo de data centers [Submarine cables and clean energy turn Ceará into a data center hub].
Sep.~1, 2025.
Available at: \url{https://www1.folha.uol.com.br/tec/2025/09/cabos-submarinos-e-energia-limpa-transformam-ceara-em-polo-de-data-centers.shtml}.

\bibitem{fgv2026}
Fundação Getulio Vargas (FGV).
\emph{Estudo de Potenciais Impactos Socioeconômicos da Consolidação do Brasil como Hub Internacional de Infraestrutura Digital na Era da Inteligência Artificial} [Study of the potential socioeconomic impacts of consolidating Brazil as an international digital infrastructure hub in the age of artificial intelligence].
FGV Projetos, 2026.
Available at: \url{https://portal.fgv.br/noticias/estudo-da-fgv-aponta-que-data-centers-com-ia-geram-mais-de-12-mil-empregos-e-mobilizam-25-bilhoes}.

\bibitem{istoe2025}
IstoÉ Dinheiro.
MPF questiona autorização dada pelo governo do Ceará para mega data center do TikTok [MPF questions the authorization granted by the Ceará government for TikTok's mega data center].
Dec.~17, 2025.
Available at: \url{https://istoedinheiro.com.br/mpf-questiona-autorizacao-dada-pelo-governo-do-ceara-para-mega-data-center-do-tiktok-2}.
Accessed: Nov.~25, 2026.

\bibitem{mpf2026}
Ministério Público Federal (MPF).
MPF e DPU vão à Justiça para impedir operação de data center no Complexo do Pecém, em Caucaia (CE) [MPF and DPU go to court to prevent the operation of a data center in the Pecém Complex, Caucaia, Ceará].
Sep.~10, 2026.
Available at: \url{https://www.mpf.mp.br/o-mpf/unidades/pr-ce/noticias/mpf-e-dpu-vao-a-justica-para-impedir-operacao-de-data-center-no-complexo-do-pecem-em-caucaia-ce}.
Accessed: Nov.~25, 2026.

\bibitem{niosh2017}
National Institute for Occupational Safety and Health (NIOSH).
New NIOSH Sound Level Meter app.
Jan.~17, 2017.
Available at: \url{https://cdc.gov/niosh/bulletin/2017/sound-app.html}.
Accessed: Nov.~25, 2026.

\bibitem{niosh_understand}
National Institute for Occupational Safety and Health (NIOSH).
Understand noise exposure.
Available at: \url{https://www.cdc.gov/niosh/noise/prevent/understand.html}.
Accessed: Nov.~25, 2026.

\bibitem{ngata2025}
W.~Ngata, N.~Bashir, M.~Westerlaken, L.~Liote, Y.~Chandio, and E.~Olivetti.
The Cloud Next Door: Investigating the environmental and socioeconomic strain of datacenters on local communities.
In \emph{Proceedings of the 2025 ACM SIGCAS/SIGCHI Conference on Computing and Sustainable Societies (COMPASS '25)}. ACM, 2025.
doi:\href{https://doi.org/10.1145/3715335.3736324}{10.1145/3715335.3736324}.
Preprint: arXiv:2506.03367.

\bibitem{ons2025}
Operador Nacional do Sistema Elétrico (ONS).
\emph{PAR/PEL 2025 -- Plano da Operação Elétrica de Médio Prazo do Sistema Interligado Nacional} [Medium-term electrical operation plan for the National Interconnected System].
2025.
Available at: \url{https://www.ons.org.br/Paginas/energia-no-futuro/suprimento-eletrico/parpel2025/sumario-executivo/index.aspx}.

\bibitem{saopaulo2016}
São Paulo (Municipality).
Lei nº~16.402, de 22 de março de 2016: disciplina o parcelamento, o uso e a ocupação do solo no Município de São Paulo [Law No.~16,402 of March 22, 2016: regulates land subdivision, use, and occupation in the Municipality of São Paulo].
Available at: \url{https://legislacao.prefeitura.sp.gov.br/lei-16402-de-22-de-marco-de-2016/consolidado}.

\end{thebibliography}
\end{document}